\documentclass[%
 aip,
 amsmath,amssymb,
reprint,%
]{revtex4-1}

\usepackage{etoolbox}
\usepackage[version=3]{mhchem} % Formula subscripts using \ce{}
\usepackage[ruled,vlined]{algorithm2e}
\usepackage{changes}
\renewcommand{\added}[1]{{\color{black}#1}}
\renewcommand{\deleted}[1]{}
\renewcommand{\replaced}[2]{{\color{black}#1}}

\makeatletter
\def\@email#1#2{%
 \endgroup
 \patchcmd{\titleblock@produce}
  {\frontmatter@RRAPformat}
  {\frontmatter@RRAPformat{\produce@RRAP{*#1\href{mailto:#2}{#2}}}\frontmatter@RRAPformat}
  {}{}
}%

\makeatother
\begin{document}

%\preprint{AIP/123-QED}
\title{Training Algorithm Matters for the Performance of Neural Network Potential: A Case Study of Adam and the Kalman Filter Optimizers}

\author{Yunqi Shao}
\affiliation{Department of Chemistry-\AA{}ngstr\"{o}m Laboratory, Uppsala University, L\"{a}gerhyddsv\"{a}gen 1, BOX 538, 75121, Uppsala, Sweden}
\author{Florian M. Dietrich}
\affiliation
{Department of Chemistry-\AA{}ngstr\"{o}m Laboratory, Uppsala University, L\"{a}gerhyddsv\"{a}gen 1, BOX 538, 75121, Uppsala, Sweden}
\author{Carl Nettelblad}
\affiliation
{Division of Scientific Computing, Department of Information Technology, Uppsala University, L\"{a}gerhyddsv\"{a}gen 2, BOX 337, 75105, Uppsala, Sweden}
\author{Chao Zhang}
 \email{chao.zhang@kemi.uu.se}
\affiliation
{Department of Chemistry-\AA{}ngstr\"{o}m Laboratory, Uppsala University, L\"{a}gerhyddsv\"{a}gen 1, BOX 538, 75121, Uppsala, Sweden}

\date{\today}

\begin{abstract}
  One hidden yet important issue for developing neural network potentials (NNPs) is the choice of
  training algorithm. Here we compare the performance of two popular training
  algorithms, the adaptive moment estimation algorithm (Adam) and the Extended Kalman Filter algorithm (EKF), using the Behler-Parrinello neural network (BPNN) and two
  publicly accessible datasets of liquid water [Proc. Natl. Acad. Sci. U.S.A.
  2016, 113, 8368-8373 and Proc. Natl. Acad. Sci. U.S.A. 2019, 116, 1110-1115]. This is achieved by
  implementing EKF in TensorFlow. It is found that  NNPs  trained  with  EKF  are  more  transferable  and  less  sensitive  to  the  value  of  the learning rate, as compared to Adam. In both cases, 
  error metrics of the \replaced{validation}{test} set do not always serve as a good indicator for the actual performance of NNPs. Instead, we show that their performance correlates well with a  Fisher information based similarity measure.
\end{abstract}

\maketitle

\section{Introduction}

Neural network potentials (NNPs) are one category of machine learning potentials~\cite{Behler21_chemrev, Deringer21_chemrev, Watanabe_2020, Unke20} which approximate potential energy surfaces (PES) and allow for large-scale simulations with the accuracy of reference electronic structure calculations but at only a fraction of the computational cost~\cite{9355242}.

One prominent architecture of NNPs are Behler-Parrinello neural networks ~\cite{2007_BehlerParrinello} which
introduced the idea of partitioning the total potential energy of the system into effective atomic contributions. BPNNs have been applied to a wide range of molecules and materials~\cite{Gastegger2017MachineSpectra, 2016_MorawietzSingraberEtAl, 2019_ChengEngelEtAl, VanessaQuaranta:2017kz, 2019_SchranBehlerEtAl, Shao:2020gp} Despite these successes, the BPNN architecture relies on the selection of a set of symmetry functions before the training in order to describe the local chemical environments.
On the contrary, features in deep-learning~\cite{Lecun2015DeepLearning} are automatically learned via hierarchical filters, rather than handcrafted. In particular, graph convolution neural networks (GCNN), which consider the atoms as nodes and the pairwise interactions as weighted edges in the graph, have emerged as a new type of architectures in constructing NNPs for both molecules and materials~\cite{Schutt2018SchNetMaterials, Chen:2018fu, 2020_ShaoHellstroemEtAl}.

Innovations regarding new architectures certainly drive the improvement of the performance of NNPs. However, one hidden and less discussed issue is the impact of training algorithms on the construction of NNPs. As a matter of fact, earlier implementations of BPNNs ~\cite{Gastegger:2015dl, Artrith:2016gc, 2018_Behler, 2019_SingraberMorawietzEtAl,Huang:2019ky} used either the Limited-memory Broyden–Fletcher–Goldfarb–Shanno algorithm (L-BFGS)~\cite{Liu:1989wl} or the Extended Kalman Filter algorithm (EKF)~\cite{Singhal88} as the default optimizer, while recent implementations of GCNN architectures~\cite{Schutt2018SchNetMaterials, Chen:2018fu, 2020_ShaoHellstroemEtAl,TorchANI20} almost exclusively chose the adaptive moment estimation algorithm (Adam)~\cite{2014_KingmaBa} instead.  This contrast is, partly, due to the convenience that efficient implementations of Adam are available on popular frameworks such as TensorFlow~\cite{tensorflow2015-whitepaper} and PyTorch~\cite{NEURIPS2019_9015}. Nevertheless, the practical equivalence among optimization algorithms for training NNPs is assumed rather than verified.

\added{In principle, comparing training algorithms for the performance of NNPs is a highly non-trivial task. This largely comes from the fact that the final performance of NNPs is affected by many different factors which could convolute the comparison. These include the construction of dataset, the setup of loss function and batch schedule, the choice of neural network architecture, the hyperparameters of training algorithm, not to mention human errors in implementing these training algorithms and the corresponding neural network architecture. Even everything is properly controlled, the ground truth to judge the effect of training algorithm on the performance of NNPs in real-application scenarios may simply not be available.  

To carry out this daunting task, our strategy is to use published datasets of liquid water where credible reference values in real-application scenarios are available~\cite{2016_MorawietzSingraberEtAl, 2019_ChengEngelEtAl}. This allows us to just focus on the effect of training algorithm, while the rest of hyperparameters, i.e. the NNP architecture (setups of the BPNN and symmetry functions), the loss function and the batch schedule, in addition to the dataset, can be inherited and kept fixed. To minimize the impact due to 
differences in implementation,
we implement EKF in the TensorFlow-based PiNN code~\cite{2020_ShaoHellstroemEtAl}, where the BPNN architecture was previously
implemented and tested. The technical soundness of our implementation of EKF is verified against the EKF implementation in the RuNNer code~\cite{2018_Behler}, which was used to generate NNPs in the reference works with the same datasets~\cite{2016_MorawietzSingraberEtAl, 2019_ChengEngelEtAl}. In this way, we can compare EKF and Adam on an equal footing, where the native implementation of Adam in Tensorflow is used. }

\deleted{In this work, by implementing EKF in TensorFlow, we systematically compare the performance of two popular training algorithms, Adam and EKF, using BPNNs and two publicly accessible datasets of liquid water. We find that training algorithms have a non-trivial impact on the performance of NNPs, both in terms of extrapolation and interpolation.} Before showing these results, we will outline the details of our comparative study, including the training algorithms, datasets, \added{the network architecture}, and molecular dynamics simulations used for the density prediction, as described in the next section. 

\section{Methods}

\subsection{\added{Loss Function and Batch Schedule}}

\added{

\begin{figure}
    \includegraphics[width=.8\linewidth]{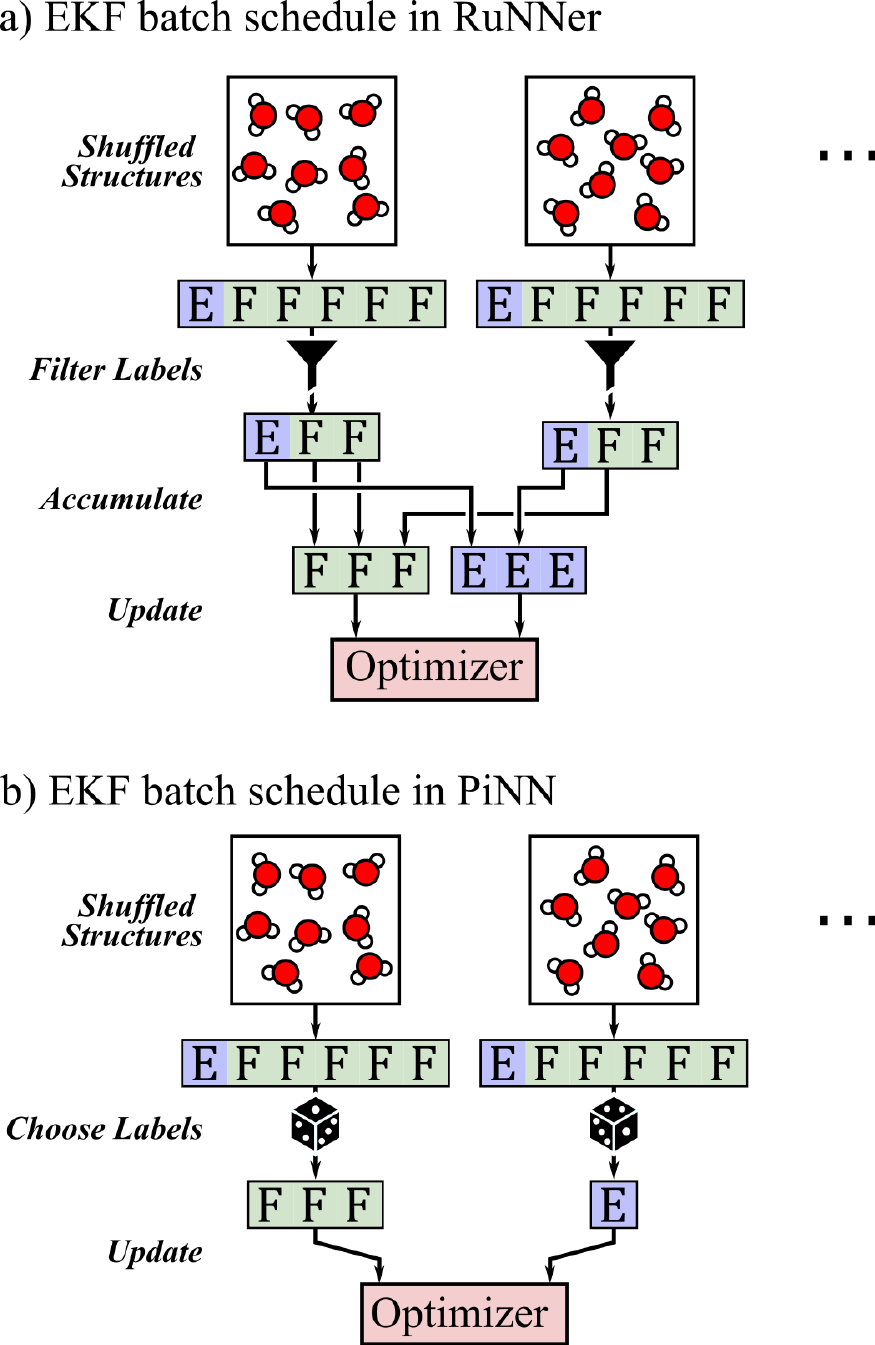}
    \caption{\added{Illustrations of the batch schedule of energy (E) and force (F) labels in a) RuNNer and b) PiNN. This is designed for EKF and used in Adam for the comparative study here.}}
    \label{fig:scheduling}
\end{figure} 

As the aim in this study is to elucidate the effect of training algorithm on the performance of NNPs ,
the definition of loss function and the choice of batch schedule need to be held consistent between EKF and Adam algorithms.

We denote $\hat{\mathbf{y}}(\mathbf{w}, \mathbf{x})$ as the neural network 
prediction given the weight vector $\mathbf{w}$ and input vector $\mathbf{x}$, and $\mathbf{y}$ as
the training-data labels.
Here, we define $\boldsymbol{\xi} = \sqrt{c/n} (\hat{\mathbf{y}}-\mathbf{y})$ as the scaled error vector, where $n$ is the size of vector $\mathbf{y}$ and $c$ is the weighting factor (for balancing number of energy and force labels). This leads to the $L_1$ loss function as $L_1 = \sum_i |\xi_i|$. The corresponding $L_2$ loss function $L_2 = \sum_i \xi_i^2$ is then related to the mean squared error scaled by $c$. Since EKF minimizes the error vector $\boldsymbol{\xi}$ and Adam minimizes the $L_2$ loss, this setup ensures that two algorithms optimize the same loss function as long as training data are the same.

Another distinction between the two algorithms (without diving into the details, as elaborated in the next Section) is on the batch schedule. EKF, as originated from the field of control theory and signal processing~\cite{Haykin:2001un}, is designed for on-line training (i.e. one training data-point at a time), in contrast to the mini-batch (a group of randomly selected training data) often used in Adam.
This difference becomes blurred when the multi-stream variant of EKF is  employed.\cite{2019_SingraberMorawietzEtAl}
Nevertheless, we followed the practice of doing weight-update step based on either energy or force as in the previous studies,
to avoid the complication of combining energy and force data in one step.\cite{2019_SingraberMorawietzEtAl}

In the RuNNer setup of EKF, the weights are updated when a given
number of error and corresponding Jacobian is computed.
As the number of force labels overwhelms that of energy labels, 
a small fraction of force labels is typically used \cite{2019_SingraberMorawietzEtAl}.
In addition, the labels can be filtered according to the magnitude of the error, 
which potentially improves the efficiency. 
An illustration of the batch schedule used in RuNNer is shown in Fig.~\ref{fig:scheduling}a.
 
Here, we choose to apply a simplified scheme of batch schedule in which the weight
update is based on randomly selected force or energy label in each iteration without filtering, which is computationally more efficient in TensorFlow. In our particular instance, it is one energy label or ten force labels for each iteration, as illustrated in Fig.~\ref{fig:scheduling}b. As shown in Result and Discussions, this simplification in batch schedule does not incur an inferior performance of the NNPs optimized with EKF. Subsequently, the same batch schedule is used in Adam for the sake of consistency.

Further details regarding the loss function and batch schedule,
including the estimated number of weight updates based on energy and forces in each scheme
and the weighting factor $c$ thereof,
are listed in the Supplementary Material (Section B).
}

\subsection{Algorithms}

In the following, we first introduce the common notations used for the optimization algorithm, and then briefly state the Adam and EKF algorithms compared within this work.

\replaced{Continuing from the previous section}{In the following}, we denote $\mathbf{g}$ as the gradient of the $L_2$ loss function with respect to the weight vector $\mathbf{w}$, 
and $\mathbf{J}$ is the Jacobian matrix of $\boldsymbol{\xi}$ with respect to $\mathbf{w}$.

The first optimization algorithm used in this work is Adam, which is a popular algorithm 
for the training of neural networks.\cite{2014_KingmaBa} The algorithm can be considered as 
an extension of stochastic gradient descent in which the first moment $\mathbf{m}(t)$ and 
second  moment $\mathbf{v}(t)$ are estimated at each step, and used as a preconditioner
to the gradients. The algorithm is shown in Algo.~\ref{alg:adam}.

\begin{algorithm}[H]
\SetKwInput{kwInit}{init}
\SetInd{0.5em}{0.5em}
\SetAlgoLined
 \kwInit{
 $t = 0$,
 $\mathbf{m}(0) = 0$,
 $\mathbf{v}(0) = 0$;
 }
 \While{not converged}{
  $t=t+1$\;
  $\mathbf{g}(t) = \nabla_\mathbf{w}L_2(t)$ \;
  $\mathbf{m}(t) = \beta_1\cdot\mathbf{m}(t-1) + (1-\beta_1)\cdot\mathbf{g}(t)$\;
  $\mathbf{v}(t) = \beta_2\cdot\mathbf{v}(t-1) + (1-\beta_2)\cdot\mathbf{g}^2(t)$\;
  $\hat{\mathbf{m}}(t) = \mathbf{m}(t)/(1-\beta_1^t)$\;
  $\hat{\mathbf{v}}(t) = \mathbf{v}(t)/(1-\beta_2^t)$\;
  $\mathbf{w}(t) = \mathbf{w}(t-1) -\eta\cdot \hat{\mathbf{m}}(t)/(\sqrt{\hat{\mathbf{v}}(t)}+\epsilon)$\;
  }
  \Return $\mathbf{w}(t)$
  \caption{The Adam optimizer, where $\eta$ is the learning rate, $\epsilon$ is
    a small number, $\beta_1$, $\beta_2$ are the exponential moving average
    factors, $\mathbf{m}$, $\mathbf{v}$ are the first and second moment
    estimates and $\hat{\mathbf{m}}(t)$ and $\hat{\mathbf{v}}(t)$ are the
    bias-corrected moment estimates. \added{Here,% the $L_2$ loss function is used, 
    $\beta_1=0.9$ and $\beta_2=0.999$ following the original publication;\cite{2014_KingmaBa}
    $\epsilon=10^{-7}/\sqrt{1-\beta_2^t}$ as the default in TensorFlow.
    }}
 \label{alg:adam}
\end{algorithm}

\begin{algorithm} [H]
\SetKwInput{kwInit}{init}
\SetInd{0.5em}{0.5em}
\SetAlgoLined
 \kwInit{
 $t = 0$,
 $\mathbf{P}(0) = \mathbf{I}$;
 }
 \While{not converged}{
  $t=t+1$\;
  $\mathbf{J}(t) = \nabla_\mathbf{w}\boldsymbol{\xi}(t)$ \;
  $\mathbf{A}(t) = \left[\mathbf{J}(t)\mathbf{P}(t-1)\mathbf{J}^\top(t)+\mathbf{R}(t)\right]^{-1}$\;
  $\mathbf{K}(t) = \mathbf{P}(t-1)\mathbf{J}^\top(t)\mathbf{A}(t)$\;
  $\mathbf{P}(t) = \left[\mathbf{I} - \mathbf{K}(t)\mathbf{J}(t)\right]\mathbf{P}(t-1)+\mathbf{Q}(t)$\;
  $\mathbf{w}(t) = \mathbf{w}(t-1) + \mathbf{K}(t)\boldsymbol{\xi}(t)$\;
  }
  \Return $\mathbf{w}(t)$
  \caption{The extended Kalman Filter (EKF) optimizer, where \added{$\boldsymbol{\xi}$ 
    is the error vector at each iteration, $\mathbf{J}$ is the corresponding 
    Jacobian matrix, } $\mathbf{A}$
    is termed as the scaling matrix, $\mathbf{K}$ is the Kalman gain,
    $\mathbf{P}$ is the error covariance matrix and \deleted{$\mathbf{R}$ is}
 the observation-noise covariance matrix \added{$\mathbf{R}$ is set to $\frac{1}{\eta} \mathbf{I}$, with $\eta$ being the learning rate.\cite{2019_SingraberMorawietzEtAl}}  \replaced{Following the setting in RuNNer~\cite{2018_Behler},}{For the sake of discussion,} the
    initial $\mathbf{P}(0)$ is set to an identity matrix and the process-noise covariance
    matrix $\mathbf{Q}$ is set to zero in this work.}
 \label{alg:ekf}
\end{algorithm}

\deleted{We denote $\hat{\mathbf{y}}(\mathbf{w}, \mathbf{x})$ as the neural network 
prediction given the weight vector $\mathbf{w}$ and input vector $\mathbf{x}$, and $\mathbf{y}$ as
the reference labels. }
\deleted{
$\boldsymbol{\xi} = \mathbf{y}-\hat{\mathbf{y}}$ is the error vector. The mean squared error $L_2 = \frac{1}{n} \sum_i \xi_i^2$ is used as the loss function
unless otherwise mentioned.}

The other optimization algorithm used in this work is EKF, which estimates
the internal state of a system given a series of observations over time. In the context
of neural network training, it can be interpreted as updating the weights of the neural network
according to the gradient of the error with respect to the weights in past
samples.\cite{Singhal88} The algorithm is summarized in Algo.~\ref{alg:ekf}.
\added{Note that the notation used here follows Ref.~\citenum{2019_SingraberMorawietzEtAl}, where the learning rate $\eta$ is controlled by the observation-noise covariance matrix $\mathbf{R}$. This formulation is more transparent as compared to earlier studies~\cite{Witkoskie:2005be}.}

\subsection{Dataset description}
Two datasets containing structures, energies and forces of liquid water (and ice phases) were used to train the NNPs in this work. 6324 structures in the BLYP dataset of both liquid phase and ice phases were taken from Ref.~\citenum{2016_MorawietzSingraberEtAl} and re-computed with the CP2K suite of programs~\cite{2020_KuehneIannuzziBenEtAl} and the BLYP
functional\cite{1988_Becke,1988_LeeYangParr}.  The revPBE0-D3 dataset of 1593 structures of liquid water computed with CP2K was directly taken from
Ref.~\citenum{2019_ChengEngelEtAl}.

\subsection{Neural Network Potential}

\subsubsection{Network architecture}

In this work, the NNPs were constructed using the BPNN architecture,\cite{2007_BehlerParrinello} with the symmetry
function taken from Ref.~\citenum{2016_MorawietzSingraberEtAl}. \added{
Specifically, 30 symmetry functions for oxygen and 27 for hydrogen were 
used for the local environment description of atoms. The exact set of symmetry 
functions used is given in the Supplementary Material (Section A).
This local description was fed to an element-specific feed-forward neural network
containing 2 hidden layers
each comprised of 25 nodes with the $\mathrm{tanh}$ activation function 
and a linear output node to give atomic energy predictions, 
from which the total energy and force predictions are derived.
The weight parameters were initialized using the Nguyen-Widrow method in RuNNer,\cite{1990_NguyenWidrow}
and the Xavier method  in PiNN.\cite{2010_GlorotBengio}
}

\deleted{
BPNN was implemented previously in the PiNN codebase~\cite{2020_ShaoHellstroemEtAl}, which is a TensorFlow-based package for building atomic neural networks for both molecules and materials~\cite{Shao2021_Batt, Knijff_2021}.
The default implementation of Adam in
TensorFlow was adopted (denoted as PiNN-Adam in the following), and
we have implemented the EFK optimizer in Tensorflow (named as PiNN-EKF). To benchmark our EKF implementation, the same
dataset was used to train a set of NNPs using the RuNNer code.\cite{2018_Behler}}

\subsubsection{Algorithm hyperparameters}

The PiNN-Adam models were trained for $5 \times 10^{6}$ steps, with a learning
rate \added{$\eta$} that decays by a factor of 0.994 every $10^{4}$ steps.
Notably, the gradient clipping technique was used with the Adam optimizer to
alleviate the vanishing and exploding gradients
problems.\cite{2013_PascanuMikolovBengio} \added{Specifically, gradient 
vectors $\mathbf{g}$ with $\|\mathbf{g}\|>0.01$ will be scaled by a factor
of $0.01/\|\mathbf{g}\|$ during training.}

The PiNN-EKF models were trained with a
learning rate \added{$\eta$} for $5 \times 10^{5}$ steps.\deleted{Learning rates were
determined with grid search, respectively for each training algorithm.} The
RuNNer models were trained for \replaced{20}{30} epochs (passes of the entire dataset), which corresponds to a total of around $3 \times 10^{5}$ steps. \added{As shown in Algo.~\ref{alg:ekf},
the learning rate $\eta$ of EKF is embedded in the covariance matrix $\mathbf{R}$ and
implicitly decays according to the inverse of iteration (see Ref.\citenum{2017_Olliviera} or Eq.~\ref{eq:ekf_weightupdate} below). 
$\eta$ is set to a constant in the PiNN setup;
the counterpart of $\eta$ asymptotically approach unity in the RuNNer setup, following the earlier studies.\cite{Witkoskie:2005be,2018_Behler}}

In both cases, 80\% of the dataset were used for training and the rest 20\% were
left out as a \replaced{validation}{test} set. For each setup, 10 models were trained with different
random seeds to split the dataset or initialize the weights of the neural
network. Whenever applicable, the standard deviations of the prediction across the models
are used as error estimates.

\subsection{Molecular Dynamics Simulation}

The molecular dynamics (MD) simulations for all the models were carried out with the ASE
code.\cite{2017_LarsenMortensenBlomqvistEtAl} The PiNN code supports calculation
with ASE, while the RuNNer code\cite{2018_Behler} was interfaced to ASE through
the LAMMPS\cite{1995_Plimpton} calculator implemented in ASE. The timestep for all simulations was
chosen to be 0.5\,fs. 
\replaced{For constant particle number, constant pressure and constant temperature (NPT) simulations, the}{The} Berendsen barostat and
thermostat\cite{1984_BerendsenPostmaGunsterenEtAl} were used to keep the pressure at 1~bar and temperature at 330 \added{or 320}~K. 
\added{For constant particle number, constant volume and constant temperature (NVT) simulations, the Berendsen thermostat
was used to keep the temperature at 300~K, and the density was fixed at 1~g/mL.}
\replaced{Each}{The} MD simulation was run for 100~ps
\added{, from which densities or radial distribution functions are computed from the later 50~ps.}
\deleted{for each of the trained models.}

\section{Result and discussions}

\subsection{The extrapolation regime: The case of the BLYP dataset}

\begin{figure} [ht]
    \centering
    \includegraphics[width=\linewidth]{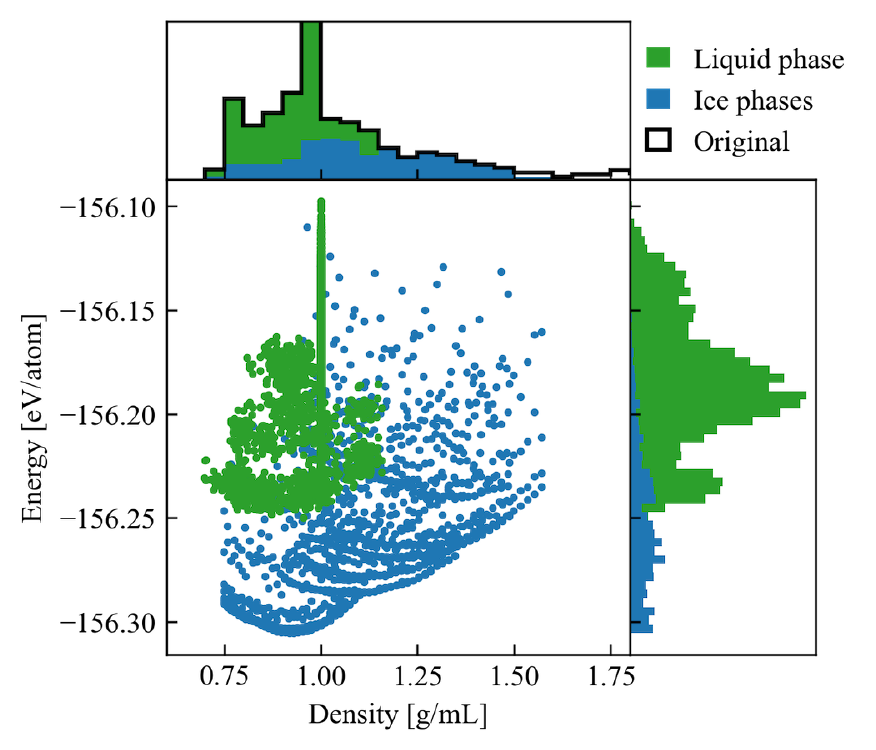}
    \caption{2D histogram of the BLYP dataset in terms of the total energy per atom and the bulk density. \added{ 
    We excluded a small fraction of the ice-phase structures
    at very high densities, as compared to the original dataset~\cite{2016_MorawietzSingraberEtAl}(shown as black outline).}}
    \label{fig:ds_stat_BLYP}
\end{figure} 

The energy-density distribution of the BLYP dataset is shown in Fig.~\ref{fig:ds_stat_BLYP}. This dataset contains structures from both liquid phase and different ice phases with a peak centering at 1.0\,g/mL.~\cite{2016_MorawietzSingraberEtAl}  Given the fact that the equilibrium density of the BLYP water at ambient conditions is below 0.8\,g/mL,~\cite{2016_MorawietzSingraberEtAl} the isobaric-isothermal density at 1\,bar and 330\,K will serve as an instructive case study where the NNP is stretched into the extrapolation regime. This motivates us to discuss the result of the BLYP dataset first, where Adam and EKF show qualitative differences. \added{Results shown in this section  were generated using the learning rate of $10^{-4}$ for PiNN-Adam and the learning rate of $10^{-3}$ for PiNN-EKF, without loss of generality (see Section C in Supplementary Material for details).}

\begin{figure}
    \centering
    \includegraphics[width=\linewidth]{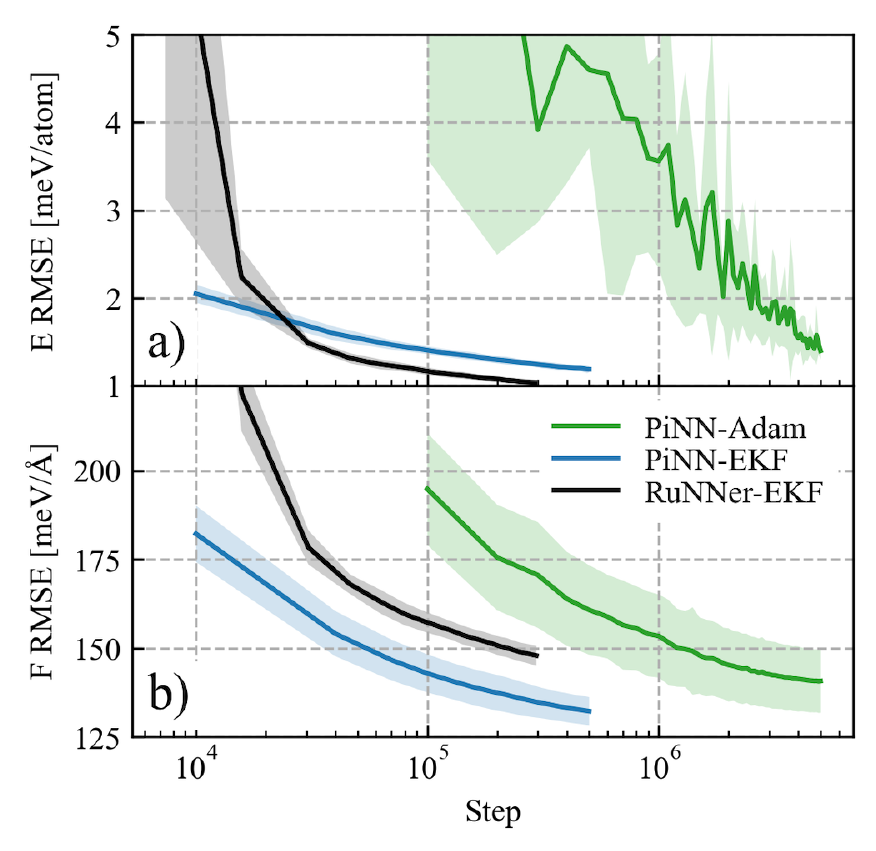}
    \caption{Training curves of the BLYP dataset for 
    a) energy root mean squared error (RMSE) and b) force RMSE
    with the Adam and the EKF optimizers.}
    \label{fig:error_log}
\end{figure}

Before presenting those results, it is necessary to first discuss the convergence \added{speed} of the NNP training. The EKF optimizer is known to converge much faster as compared to Adam (by approximately one order of
magnitude in terms of the number of weight updates to achieve the desired
accuracy)~\cite{1992_RuckRogersEtAl,2019_SingraberMorawietzEtAl}. This phenomenon is clear in our training of NNPs, as shown in Fig.~\ref{fig:error_log}. However, the actual speed-up is compromised due to the higher computational
cost of EKF. In practice, the training takes about 2\,h for the EKF optimizer and about 5\,h
for the Adam optimizer on a \replaced{28}{16}-core computer to achieve
similar levels of accuracy in terms of force and energy. It should be noted that the relative speed advantage per iteration increases drastically when the total number of weights grow.

\begin{figure}
    \centering
    \includegraphics[width=\linewidth]{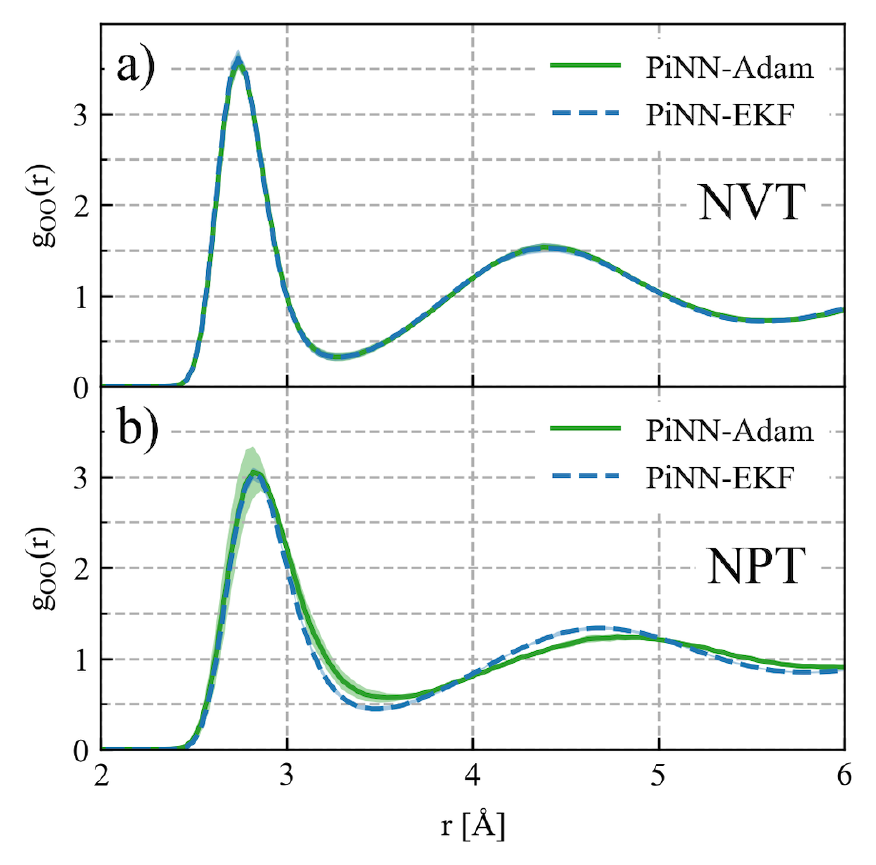}
    \caption{\added{O-O radial distribution function $g_\mathrm{OO}(r)$ for 
    a) NVT simulation at 300K, 1g/mL  using potentials trained with the Adam 
    and the EKF optimizers; 
    b) NPT simulation at 330K, 1bar using the same potentials.
    The standard deviations between models are shown as transparent regions, 
    which are small in most cases, except NPT results from PiNN-Adam.}}
    \label{fig:blyp_rdf}
\end{figure}

\added{
When it comes to the performance of NNPs, one would expect that all the trained potentials with comparable error metrics should yield a consistent water structure at room temperature and experimental density, given the dataset shown in Fig.~\ref{fig:ds_stat_BLYP}. Indeed, this is borne out,  as 
demonstrated with the O-O radial distribution function in Fig~\ref{fig:blyp_rdf}a. However, this agreement diverges quickly when it comes to isobaric-isothermal simulations using the same NNPs, as shown in Fig.~\ref{fig:blyp_rdf}b. This suggests the corresponding densities should also differ from each other.}

\begin{figure}
    \centering
    \includegraphics[width=\linewidth]{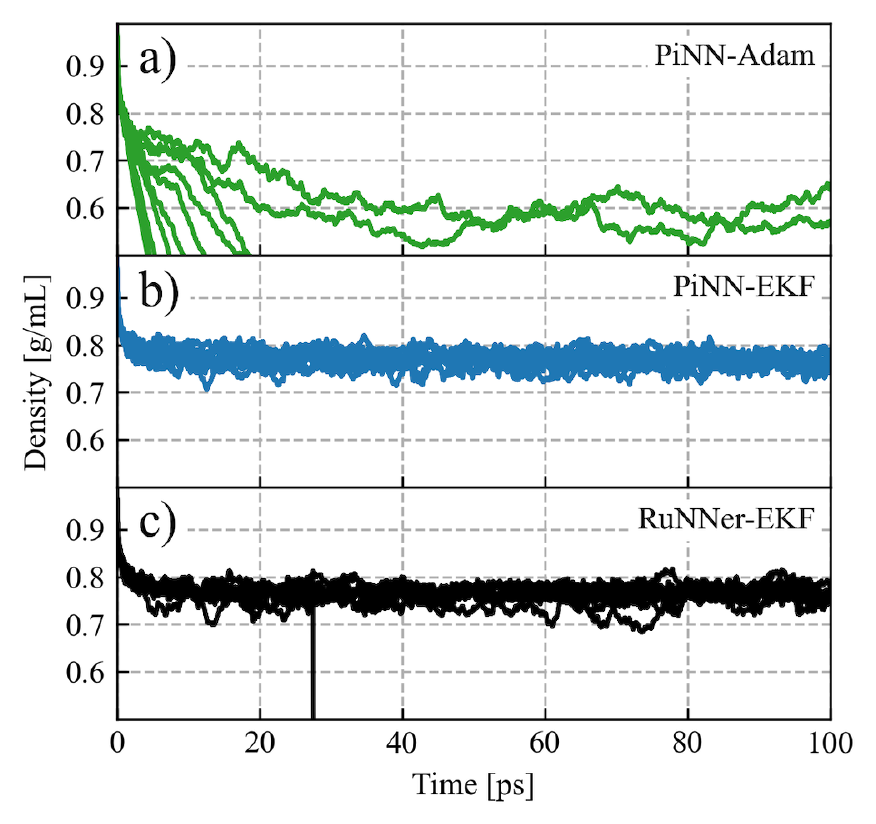}
    \caption{The density \added{evolution in} NPT simulations using potentials trained with \replaced{a) the Adam, b) and c) the EKF optimizers at 330K and 1bar}{different algorithms}.}
    \label{fig:md_density}
\end{figure}

\replaced{We then proceed}{Now we are ready}  to see how well Adam and EKF estimate the isobaric-isothermal density at ambient conditions for the BLYP dataset. As shown in Fig~\ref{fig:md_density}, at the pressure of
1\,bar and the temperature of 330\,K, only 2 out of 10 models trained with Adam manage to predict a stable
density, while most of the models trained using EKF lead to an excellent agreement with the
previously reported density of around 0.76\,g/mL for BLYP water.~\cite{2016_MorawietzSingraberEtAl} In addition, the EKF implementation in PiNN can reproduce the results of RuNNer well for the same dataset. 

The qualitative difference between NNPs trained with Adam and EKF seen in Fig.~\ref{fig:md_density} is striking, in light of comparable force and energy errors shown in Fig.~\ref{fig:error_log} \added{and radial distribution functions shown in Fig.~\ref{fig:blyp_rdf}a}. One may argue that the extrapolation is a difficult task for NNPs, because it requires stepping out from their comfort zone. Thus, it is also relevant to consider how the two training algorithms would fare within an interpolation regime. This leads to our second example, the \replaced{r}{R}evPBE0-D3 dataset.

\subsection{The interpolation regime: the case of the \replaced{r}{R}evPBE0-D3 dataset}

\begin{figure} [h]
    \centering
    \includegraphics[width=.5\textwidth]{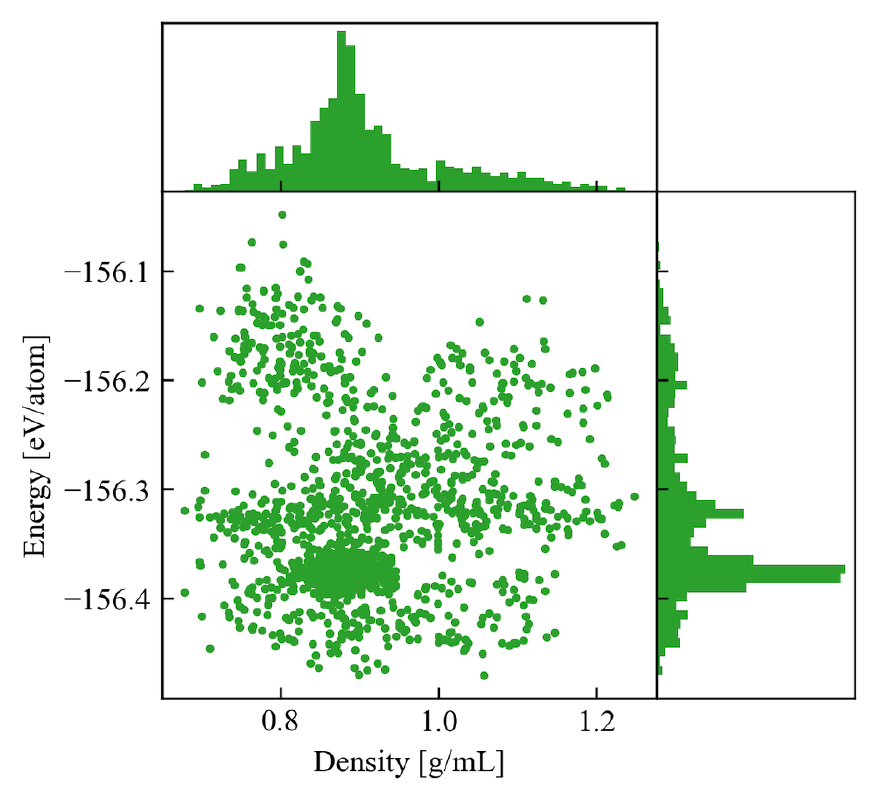}
    \caption{2D histogram of the revPBE0-D3 dataset\added{\cite{2019_ChengEngelEtAl}} in terms of the total energy per atom and the bulk density.}
    \label{fig:ds_stat_revPBE0}
\end{figure}

In the following experiment, we have used a dataset that was constructed  
to cover a wide range in the configuration space uniformly,~\cite{2019_ChengEngelEtAl} as seen in the energy-density distribution in Fig.~\ref{fig:ds_stat_revPBE0}. 

\begin{figure} [h]
    \centering
    \includegraphics[width=\linewidth]{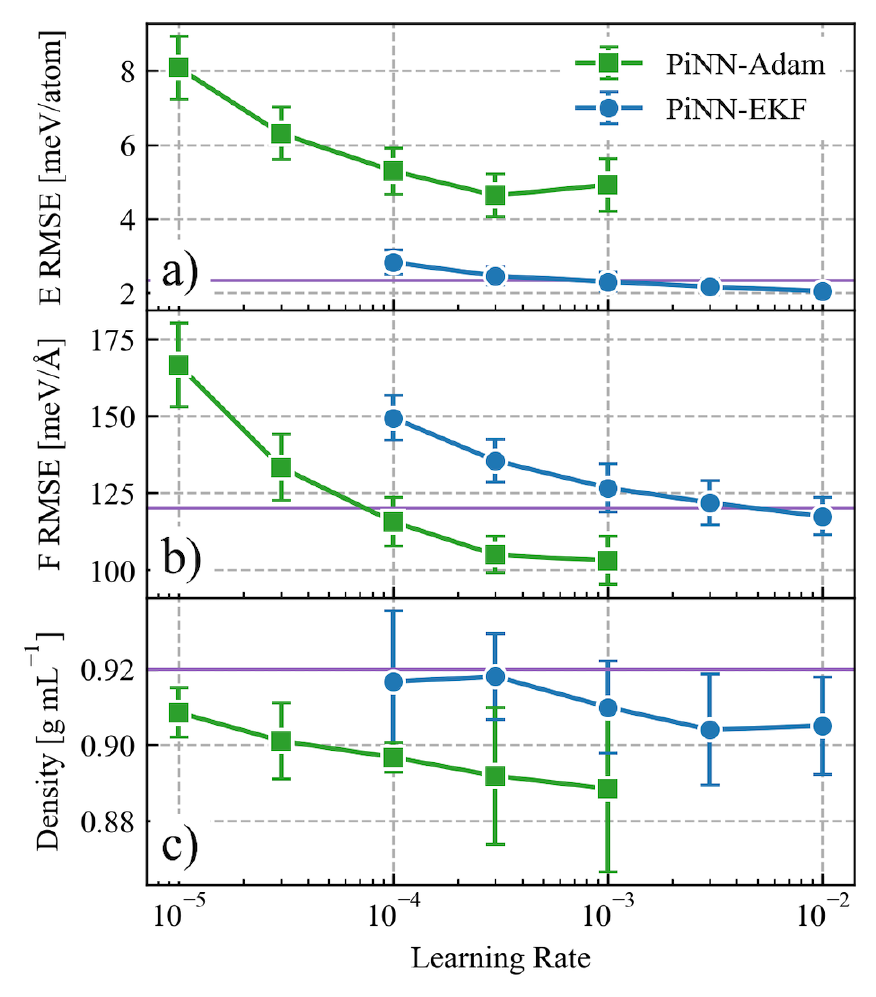}
    \caption{RMSE and density predictions for NNPs trained with 
    \replaced{the Adam and the EKF}{different} algorithms \added{at different learning rates}: a) energy RMSE; b) force RMSE and
    c) density predictions at 320K, 1bar. Reference values from 
    ref.~\citenum{2019_ChengEngelEtAl} is shown in purple.}
    \label{fig:revpbe0_lr_density}
\end{figure}

With this dataset, both Adam and EKF yield physical densities at the given temperature. \added{All of 10 models trained with Adam, regardless of the learning rate, lead to stable NPT simulations, in stark contrast to the case using BLYP dataset (see Fig. S1 in the Supplementary Material for comparison).}  However, as shown in Fig.~\ref{fig:revpbe0_lr_density}, the density prediction strongly depends on the learning rate used to train the model in the case of Adam.

It is tempting to conclude that the models trained with larger
learning rates are the better performing ones due to smaller force and energy errors. Similarly to the observation made for the BLYP dataset, the error metrics do not seem to correspond to the actual prediction performance of the trained NNPs. Indeed, we notice that 
the density does not seem to converge when the error metrics (Fig.~\ref{fig:revpbe0_lr_density}) have reached lower values. Given that the reference density value for this dataset at 1\,bar and 320\,K is 0.92\,g/ml~\cite{2019_ChengEngelEtAl}, this suggests that one has to rethink the common wisdom on performing model selection based on error metrics, \added{such as RMSE}.\added{\cite{Fonseca:2021ko,2021_KovacsOordEtAl}
}

Regardless of this implication, the 
present case study demonstrates that the performance of NNPs can be sensitive to the learning rate, in particular for Adam. As a consequence, this can lead to a tangible difference in terms of the density prediction even
within the interpolation regime where the dataset already has a good coverage. 

\subsection{Differences in models trained using Adam and EKF}  

In the previous sections, we found that NNPs trained with EKF seem to be less sensitive to the learning rate and more generalizable. Meanwhile, it is also clear that model selection based on the error metrics of energy and force may not always lead to a sound choice. 
\added{While unintuitive, the observation is not completely unexpected. 
The error metrics are typically evaluated on randomly left-out validation sets,
which underestimates the generalization error due to their correlation with the training set.
Since the training set has to cover sufficient region in the configuration space, 
it is hard to construct a test set independent of the training set.
}

Then, the questions are i) can we distinguish ``good'' and ``bad'' NNPs by just looking into the weight space (rather than doing the actual MD simulations for the density prediction)? ii) why does EKF do better for training NNPs than Adam?

To answer the first question, we have analysed the models presented in the earlier sections to shed some light on the characteristics of the models trained 
with Adam and EKF. 

One \replaced{possible}{common} measure to distinguish different models is the Euclidean distance of the optimized weights to those in the initialization
\added{, which characterizes the implicit regularization effect of stochastic gradient descent}.\cite{2019_NagarajanKolter}
Denoting the vector of all weight variables in the neural network as
$\mathbf{w}$, we \deleted{also} compared the evolution of the weights distance from
initialization $\|\Delta\mathbf{w}\|=\|\mathbf{w}-\mathbf{w}_\mathrm{init}\|$,
and \added{also} the \added{initialization-independent}
distribution of the weights $w_i$ of the final models for the trained networks.

\begin{figure}
    \centering
    \includegraphics[width=\linewidth]{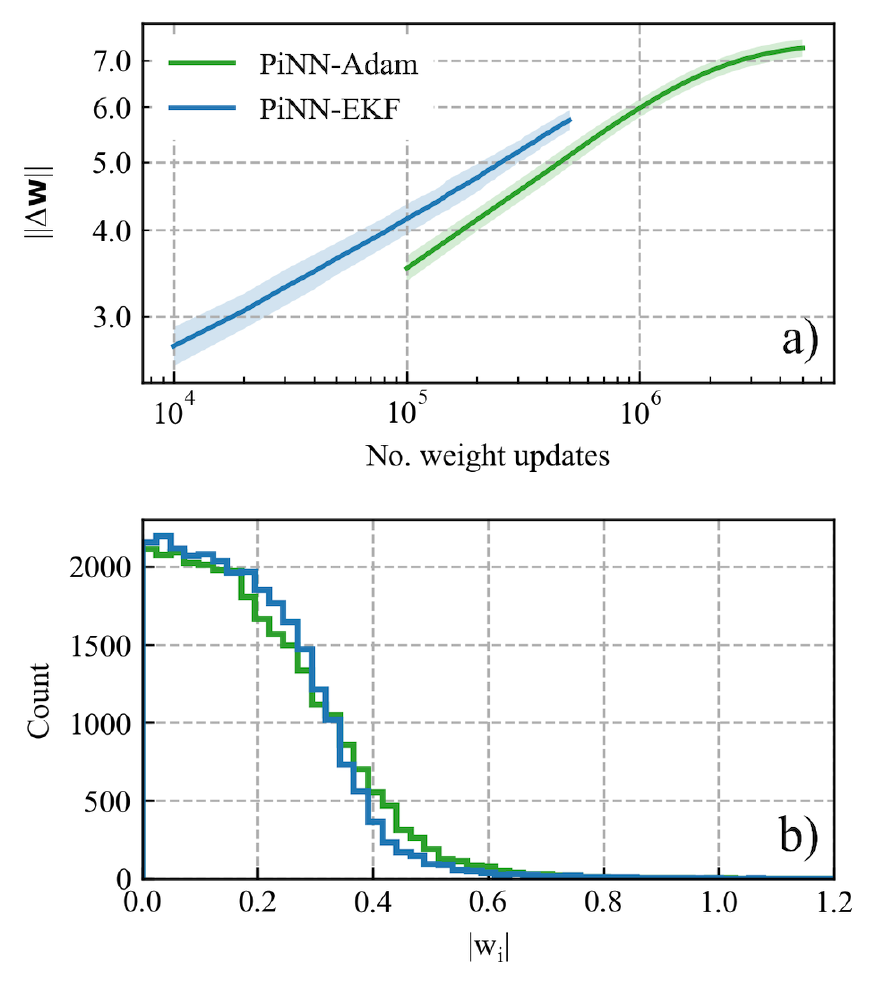}
    \caption{Weight-based measures for NNPs trained with Adam and EKF for the BLYP dataset: a) the evolution of the Euclidean weight distance from initialization
    $\|\Delta\mathbf{w}\|$ and b) the \added{initialization-independent} distribution of weights $w_i$.}
  \label{fig:weight_dist}
\end{figure}

As shown in Fig~\ref{fig:weight_dist}a, different NNPs trained on the BLYP dataset display
similar weights distances from initialization. Moreover, the weight distributions of
NNPs trained with Adam and EKF are almost indistinguishable as shown in
Fig.~\ref{fig:weight_dist}b. This suggests that the norm-based measure also
fails to distinguish models obtained from Adam and EKF. 
It further hints that using a $L_2$ norm-based regularization may not improve the performance of NNPs.

\begin{figure}
    \centering
    \includegraphics[width=\linewidth]{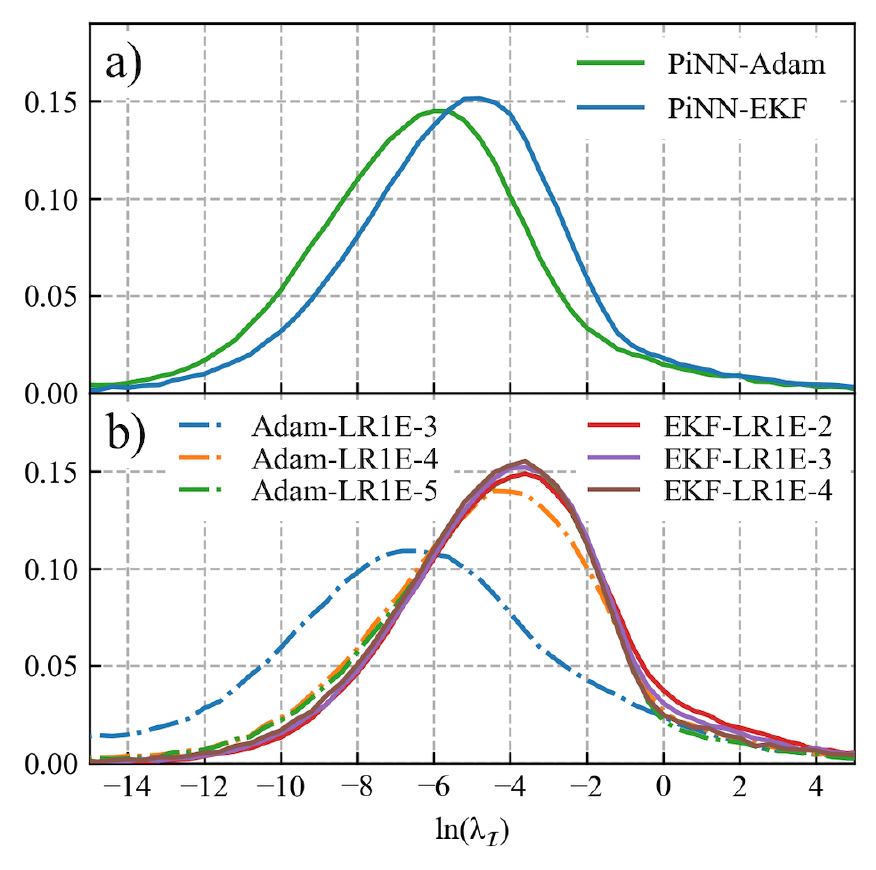}
    \caption{Distributions of the eigenvalues ($\lambda_{\mathcal{I}}$) of the Fisher information matrix:
    a) with different training algorithms for the BLYP dataset; b) with different training algorithm
    and learning rates \added{(LR)} for the revPBE0-D3 dataset.}
    \label{fig:eigen}
\end{figure}

Another class of similarity measure is related to the local information 
geometry of the neural network.\cite{2017_LiangPoggioEtAl}
Here, we characterized it with the Fisher information matrix $\mathcal{I}$\replaced{:}{.} 

\begin{equation}
    \label{eq:fisher}
\langle\mathcal{I}_1\rangle_t = \left\langle \left(\nabla_{\mathbf{w}}L_1(t)\right)^{\otimes2} \right\rangle 
\end{equation}

\added{where $^{\otimes2}$ denotes the 2nd tensor power of the gradient vector.} 
Here we used $L_1$ loss instead of $L_2$ loss to construct the Fisher information matrix for reasons that will be clear later. Fig.~\ref{fig:eigen} shows the 
distribution of eigenvalues $\lambda_{\mathcal{I}}$ of the Fisher information.
For the BLYP dataset, EKF found local minima with much larger
eigenvalues as compared to Adam, where the peak of the distribution differs by about one order of magnitude (note the logarithm 
scale used here). Therefore, the Fisher information is a similarity measure which can effectively distinguish the performance of NNPs. 

A similar trend was observed for the revPBE0-D3 dataset. The eigenvalue distribution of NNP trained with Adam and a small learning rate (1E-5) is more close to those trained with EKF. This deviation becomes larger when increasing the learning rate (1E-4 and 1E-3). In contrast, eigenvalue distributions of NNPs trained with EKF are almost identical. These results correlate well with the density prediction from MD simulations as shown in Fig.~\ref{fig:revpbe0_lr_density}.

The usefulness of the Fisher information for model selection is also linked to the second question posed at the beginning of this section.  Despite different motivations, both Adam and EKF can be understood in terms
of natural gradient descent (NGD), where the inverse of the Fisher
information matrix $\mathcal{I}$ is used as a preconditioner.

In Adam, $\mathcal{I}$ is approximated as a diagonal matrix, with the diagonal terms
thereof being the second moment vector $\hat{\mathbf{v}}(t)$. The inverse square
root of the diagonal elements are used as a conservative preconditioner.

On the other hand, the connection between EKF and NGD has been shown recently~\cite{2017_Olliviera}, where the inverse $\mathcal{I}$ matrix is
effectively estimated by the error covariance matrix $\mathbf{P}$ in EKF. In fact, one can show that
\begin{equation}
    \label{eq:ekf_weightupdate}
    \mathbf{w}(t) = \mathbf{w}(t-1) - \frac{\eta^2}{2t}\langle\mathcal{I}_1\rangle^{-1}_t\mathbf{g}(t)
\end{equation}

\replaced{when}{where} the covariance matrix of observation noise $\mathbf{R}$ in Algo.~\ref{alg:ekf} is chosen to be a scaled identity
matrix~\cite{1992_RuckRogersEtAl,2019_SingraberMorawietzEtAl,2017_Olliviera}, i.e. $\mathbf{R}(t)=\frac{1}{\eta}\mathbf{I}$.

This means EKF can be viewed as the online estimate of the full Fisher information $\mathcal{I}$ 
with respect to $L_1$ loss and the $1/t$ decay of the learning rate. We think such feature leads to a good similarly measure of models using the Fisher information based on Eq.~\ref{eq:fisher} and a superior performance of EKF for training NNPs as demonstrated in two case studies shown in this work.

\section{Conclusion and outlook}

To sum up, we have compared the performance of two optimization algorithms Adam and EKF for training NNPs of liquid water based on BLYP and revPBE0-D3 datasets. It is found that NNPs trained with EKF are more transferable and less sensitive to the choice of the learning rate, as compared to Adam. Further, we show that the Fisher information is a useful similarity measure for the model selection of NNPs when the error metrics and the normal-based measure become ineffective.

Established practice in other neural network applications~\cite{pmlr-v37-ioffe15}, using Adam, indicate that practical training performance can be improved if an architecture is reparametrized to promote weights in a common numerical range, without strong correlations.  Another future avenue would thus be to preserve the expressive power of existing NNPs, while expressing the weights in such a way that the off-diagonal elements of the resulting Fisher information matrix are minimized. This can be understood in terms of the limitations in the Adam preconditioner strategy.

\added{On the other hand, EKF scales quadratically with the number of trainable parameters in the network, while Adam scales linearly.
This makes EKF a less favorable choice when more heavily parameterized models like GCNN\cite{Schutt2018SchNetMaterials, Chen:2018fu, 2020_ShaoHellstroemEtAl} are of interest. 
Necessary approximation to the estimation of $\mathcal{I}$, 
such as that based on Kronecker factorization,\cite{2015_MartensGrosse} 
is essential to transfer the present observation to those models.}

Before closing, it is worth to note that the present issue of training algorithms may be overcome with more data \added{(as shown in Fig.~9 with augmented dataset in our previous work~\cite{2020_ShaoHellstroemEtAl}), for which a number of active-learning approaches tailored for NNPs have been proposed.\cite{2012_ArtrithBehler,Smith:2018gja, 2019_SchranBehlerEtAl,2020_ZhangWangEtAl,2021_SchranThiemannEtAl}} However, we would argue that a better training algorithm does not only improve the performance of NNPs but also improve the data efficiency in the active-learning. 

\added{This work focuses on the practical role of training algorithm in the performance of NNPs, where different approximations to the Fisher information (in Adam and EKF) and the choice of learning rate are factors primarily considered. Nevertheless, as mentioned at the very beginning, other factors, such as loss function and batch schedule, neural network architecture, decay factor and decay steps in training algorithm will bring new dimensions into the discussion. In addition, the examples shown here are based on published dataset of liquid water and the real-application scenario is density prediction. Therefore, the conclusion drawn in this work is yet to be confirmed in more complex systems and other physico-chemical properties, where credible reference values in real-application scenarios can be obtained. Overall, this calls for more systematic investigations to shed light on this topic and the practice of open science to facilitate similar studies in the community. With these efforts, the development of NNPs will become more transparent and reproducible, which in turn will provide reliable insights into interesting chemical physics and physical chemistry problems.}

\section*{Supplementary Material}

See Supplementary Material for symmetry function parameters, weighting
factor in batch schedule and loss Function, and learning rate dependency of error metrics and density
for the BLYP dataset.

\begin{acknowledgements}
 CZ thanks the Swedish Research Council (VR) for a starting grant (No. 2019-05012). We also acknowledge the Swedish National Strategic e-Science program eSSENCE for funding. Part of the simulations were performed on the resources provided by the Swedish National Infrastructure for Computing (SNIC) at UPPMAX and NSC.
\end{acknowledgements}

\section*{Data Availability Statement}
Datasets used in this work are publicly accessible from
DOI: 10.5281/zenodo.2634097 and
DOI: 10.24435/materialscloud:2018.0020/v1. The PiNN code is
available from https://github.com/Teoroo-CMC/PiNN and its EKF
implementation will be released in the next version and provided upon
request in the interim.

\end{document}

\newpage 
\section{A. Symmetry Function Parameters}

In this work, atomistic descriptions are given with 
the $G^2$ and $G^3$ symmetry functions\cite{2018_Behler}:

\begin{align}
    G^2_i &= \sum_j e^{-\eta (r_{ij} - r_s)^2} \cdot f_c(r_{ij})\\
    G^3_i &= 2^{1-\zeta}\sum_{j\ne i}\sum_{k\ne i,j}[
    (1+\lambda\cdot\cos{\theta_{ijk}})^\zeta
    \cdot e^{-\eta (r_{ij}^2 + r_{ik}^2 + r_{jk}^2)}\notag \\ &\qquad
    \cdot f_c(r_{ij})\cdot f_c(r_{ik})\cdot f_c(r_{jk})] \\
    f_c(r_{ij}) &= \mathrm{tanh}^3\left(1-r_{ij}/r_c\right)
\end{align}

where $f_c(r)$ is the cutoff function; $r_c=12\ \mathrm{Bohr}$ for all symmetry functions; and
$\eta$, $r_s$ and $\lambda$ are parameters that specify the shape of the symmetry functions.
The parameters for the $G^2$ and $G^3$ symmetry functions used in this work is listed in 
Table.~S\ref{tab:sf_radial} and S\ref{tab:sf_angular}.
The symmetry functions are then scaled to the interval $[-1,1]$ according to their
range in the training set, by applying:

\begin{equation}
    G_i^\mathrm{scaled} = \frac{2(G_i-G_{i,\mathrm{min}})}{G_{i,\mathrm{max}}-G_{i,\mathrm{min}}}-1
\end{equation}

\begin{table}
    \centering
    \begin{tabular}{ccrr}
    \hline
    \hline
     Element i        &  Element j         & $r_s$ [$\mathrm{Bohr}$] &  $\eta$ [$\mathrm{Bohr}^{-2}$]  \\
    \hline
    H         & H          & 0.0                     &  0.001, 0.01, 0.03, 0.06 \\
    H         & H          & 1.9                     &  0.15, 0.30, 0.60, 1.50  \\
    H         & O          & 0.0                     &  0.001, 0.01, 0.03, 0.06 \\
    H         & O          & 0.9                     &  0.15, 0.30, 0.60, 1.50  \\
    O         & H          & 0.0                     &  0.001, 0.01, 0.03, 0.06 \\
    O         & H          & 0.9                     &  0.15, 0.30, 0.60, 1.50  \\
    O         & O          & 0.0                     &  0.001, 0.01, 0.03, 0.06 \\
    O         & O          & 4.0                     &  0.15, 0.30, 0.60, 1.50  \\
    \hline
    \hline
    \end{tabular}
    \caption{Definition of $G^2$ symmetry in the description, for each combination of 
    element pair and $r_s$, a set of symmetry functions is used whose $\eta$ 
    parameters are enumerated. }
    \label{tab:sf_radial}
\end{table}

\begin{table}
    \centering
    \begin{tabular}{cccrrr}
    \hline
    \hline
    Element i &  Element j & Element k & $\lambda$ &  $\zeta$ & $\eta$ [$\mathrm{Bohr}^{-2}$]         \\
    \hline
    H  & H & O &  1.0 &  4.0, 1.0, 1.0, 1.0 &  0.01, 0.03, 0.07, 0.20\\
    H  & H & O & -1.0 &  4.0, 1.0, 1.0      &  0.01, 0.03, 0.07      \\
    H  & O & O &  1.0 &  4.0, 1.0           &  0.001, 0.03           \\
    H  & O & O & -1.0 &  4.0, 1.0           &  0.001, 0.03           \\
    O  & H & H &  1.0 &  4.0, 1.0, 1.0      &  0.01, 0.03, 0.07      \\
    O  & H & H & -1.0 &  4.0, 1.0, 1.0      &  0.01, 0.03, 0.07      \\
    O  & H & O &  1.0 &  4.0, 1.0           &  0.001, 0.03           \\
    O  & H & O & -1.0 &  4.0, 1.0           &  0.001, 0.03           \\
    O  & H & H &  1.0 &  4.0, 1.0           &  0.001, 0.03           \\
    O  & H & H & -1.0 &  4.0, 1.0           &  0.001, 0.03           \\
    \hline
    \hline
    \end{tabular}
    \caption{Definition of $G^3$ symmetry in the description, for each combination of 
    element triplet and $\lambda$, a set of symmetry functions is used whose $\eta$ and $\zeta$ 
    parameters are enumerated. }
    \label{tab:sf_angular}
\end{table}

\section{B. Weighting Factor in Batch Schedule and Loss Function}

The difference in the scheduling of labels means the frequency and weighing factor
of weight updates with energy and force labels in RuNNer and PiNN are different. 
However, the error vector and the loss function used are consistent between PiNN-Adam and PiNN-EKF.
The approximated number of updates for each type of label per epoch, and the weighting-factor ratio for energy and force losses for both setups are listed in Table.~S\ref{tab:scheduling}.

\begin{table}
    \centering
    \begin{tabular}{lrrr}
    \hline\hline
            & E updates & F updates & $c_\mathrm{E}$/$c_\mathrm{F}$ [a.u.]\\
    \hline
    PiNN   &   $\sim$2500       & $\sim$2500        &   0.28  \\
    RuNNer &   $\sim$900--3000  & $\sim$6000--14000 &   3.69--14.75 \\
    \hline\hline
    \end{tabular}
    \caption{Estimated number of steps used per epoch and weighing factor for
    energy (E) and force (F) labels in RuNNer and PiNN batch schedules. This estimation
    is obtained according to the statistics gathered when training on the BLYP dataset.
    The weighting factors used in the loss function are given in atomic units.
    Note that in the RuNNer batch schedule, the update refers to that based on a single label,
    which are subsequently applied in groups of 10 labels. 
    The labels are filtered according to relative error in the RuNNer scheduling,  
    therefore the number of updates are given in as ranges. 
    The relative weighing factor $c_\mathrm{E}$/$c_\mathrm{F}$ in the RuNNer setup is automatically 
    adjusted according to the number of force and energy labels in each structure. Here, 
    typical values (corresponding to structures with 16 water molecules) are given.
    }
    \label{tab:scheduling}
\end{table}

\section{C. Learning rate dependency of error metrics and density for the BLYP dataset}

\begin{figure} [h]
    \centering
    \includegraphics[width=.5\textwidth]{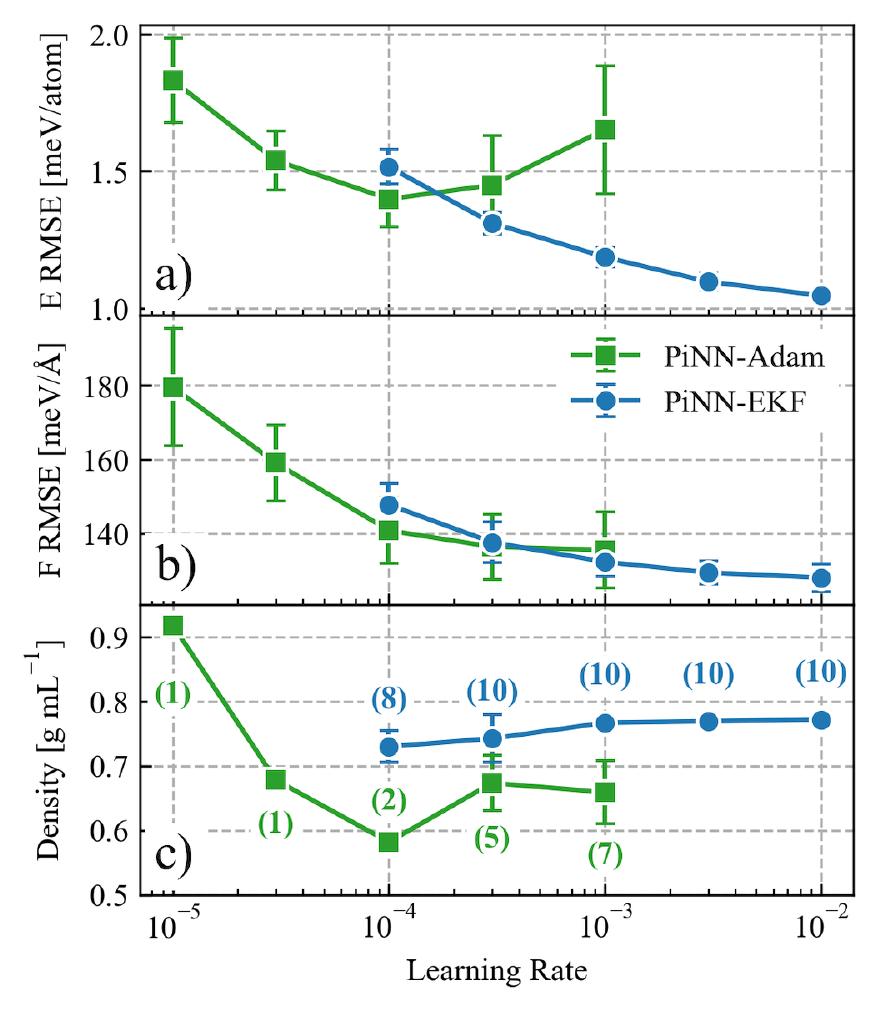}
    \caption{RMSE and density predictions for NNPs trained with 
    the Adam and the EKF algorithms at different learning rates: a) energy RMSE; b) force RMSE and
    c) density predictions at 330K, 1bar.
    The error bars are estimated using the standard deviation between models trained with 
    the same setup whenever possible. The number of models (out of 10 instances) that predict stable densities
    is annotated in the parentheses in panel c.}
    \label{fig:hdnnp_lr_density}
\end{figure}

The effect of learning rate on the error metrics and density prediction for the BLYP dataset is shown
in Fig.~S\ref{fig:hdnnp_lr_density}. In the Main Text, the learning rate for Adam is set to $10^{-4}$
to represent the best hyperparameter determined by the error metrics; the learning rate for EKF
is set to $10^{-3}$.

\providecommand{\latin}[1]{#1}
\makeatletter
\providecommand{\doi}
  {\begingroup\let\do\@makeother\dospecials
  \catcode`\{=1 \catcode`\}=2 \doi@aux}
\providecommand{\doi@aux}[1]{\endgroup\texttt{#1}}
\makeatother
\providecommand*\mcitethebibliography{\thebibliography}
\csname @ifundefined\endcsname{endmcitethebibliography}
  {\let\endmcitethebibliography\endthebibliography}{}